\documentclass[twocolumn,pra,showpacs,preprintnumbers,amsmath,amssymb]{revtex4}

\usepackage{graphicx}
\usepackage{dcolumn}
\usepackage{bm}

\begin{document}

\preprint{APS/123-QED}

\title{Countersuperflow instability\\
 in miscible two-component Bose--Einstein condensates}

\author{Shungo Ishino and Makoto Tsubota}
\affiliation{%
Department of Physics, Osaka City University, Sumiyoshi-ku, Osaka 558-8585, Japan
}%

\author{Hiromitsu Takeuchi}
\homepage{http://hiromitsu-takeuchi.appspot.com/}
\affiliation{Graduate School of Integrated Arts and Sciences, Hiroshima
University, Kagamiyama 1-7-1, Higashi-Hiroshima 739-8521, Japan}
\email{hiromitu@hiroshima-u.ac.jp}

\date{\today}

\begin{abstract}
 We study theoretically the instability of countersuperflow, {\it i.e.}, two counterpropagating miscible superflows, in uniform two-component Bose-Einstein condensates.
 Countersuperflow instability causes mutual friction between the superfluids, causing a momentum exchange between the two condensates, when the relative velocity of the counterflow exceeds a critical value.
 The momentum exchange leads to nucleation of vortex rings from characteristic density patterns due to the nonlinear development of the instability.
 Expansion of the vortex rings drastically accelerates the momentum exchange,
 leading to a highly nonlinear regime caused by intervortex interaction and vortex reconnection between the rings.
 For a sufficiently large interaction between the two components, rapid expansion of the vortex rings causes isotropic turbulence and the global relative motion of the two condensates relaxes.   
 The maximum vortex line density in the turbulence is proportional to the square of the relative velocity. 
\end{abstract}


\maketitle

\section{Introduction}
 Hydrodynamic instability, which causes exotic patterns or turbulence in the nonlinear development of complex flow structures, occurs universally throughout nature from the subatomic to the cosmic scale and has been actively studied in many fields such as magnetohydrodynamics, plasma physics, elasticity, rheology, and general relativity \cite{Kund}.
 Such phenomena can occur in quantum fluids as well as classical fluids.
 Quantum turbulence, {\it i.e.}, turbulent states in superfluids, has been thoroughly studied mainly in superfluid helium systems and has attracted considerable attention as an idealized prototype of classical turbulence \cite{Halperin2009TPLTP}, which is the most important unsolved problem of classical physics.
 In quantum fluids of atomic Bose-Einstein condensates (BECs), there is growing interest in the study of the counterparts to hydrodynamic phenomena in classical fluid dynamics, such as Kelvin--Helmholtz instability \cite{TakeuchiSKST2010PRB81,SuzukiTKTS2010PRA82}, Rayleigh-Taylor instability \cite{SasakiSAS2009PRA80,GautamA2010PRA81},
 the B$\acute{\rm e}$nard-von K$\acute{\rm a}$rm$\acute{\rm a}$n vortex street \cite{SasakiSS2010PRL104}, and turbulence \cite{KobayashiT2007PRA76,HennSRMB2009PRL103,WhiteBPYW2010PRL104}. 
 Of greatest importance in studying hydrodynamic phenomena in this system is using highly developed experimental techniques, that allow direct observation of a variety of exotic nonlinear dynamics caused by macroscopic quantum effects, {\it i.e.}, superfluidity and quantized vortices.

 Superfluid systems may also exhibit unique instability phenomena, which never appear in classical fluid dynamics.
 One such phenomenon is the instability of counterflow, in which two counterpropagating fluid components merge into each other.
 Such a counterflow state is difficult to obtain in classical fluid systems because of the viscosity between the two components,
 but is possible in superfluid systems where frictionless flow is possible in thermal equilibrium.
 In general, counterflow states become unstable for a large relative velocity between the two components.

  The instability of counterflow is classified into two cases.
 In the first case, the instability occurs between a viscid normal fluid component and an inviscid superfluid component.
 In the second case, the instability is between two superfluid components.
 The former instability has in the past been studied in superfluid ${}^4$He \cite{DonnellyBook1991},
 where the hydrodynamics is usually described using a two-fluid model in which the
system consists of a viscous normal fluid component and an inviscid superfluid component \cite{KhalatnikovBook1965}.
 The counterflow states of the two components are realized by an injected heat current, where the two components flow in opposite directions to reduce the temperature gradient.
 When the relative velocity between the two components exceeds a critical value, the thermal counterflow becomes unstable, developing into quantum turbulence with a tangle of quantized vortices.
 In the transition to the turbulent states, remnant vortices, {\it i.e.}, quantized vortices attached to the container wall, are stretched by the mutual friction between the superfluid and normal fluid components, growing into a tangle through reconnections with other vortices.
 Counterflow turbulence was visualized only recently in experiments \cite{PaolettiFSL2008JPSJ77} and is providing an active ground of study in quantum turbulence \cite{AdachiFT2010PRB81,SciaccaSBS2010PRB82,GuoCNVM2010PRL105}.

 It is interesting that counterflow instability causes mutual friction, {\it i.e.}, decay of relative motion,
 even in countersuperflow, which is counterflow between two superfluids without viscid normal fluids.
 Recently, we have reported that the instability of countersuperflow, namely, countersuperflow instability (CSI), can also develop into quantum turbulence in gaseous two-component BECs \cite{TakeuchiIT2010PRL105}.
 CSI has been studied theoretically in several multicomponent miscible superfluid systems, including helium superfluids \cite{Khalatnikov2,Mineev,Nepomnyashchii, Yukalov},
 mixture BECs of cold atoms \cite{Law}, and nucleon superfluids in rotating neutron stars \cite{Andersson}.
 Although CSI had been studied only theoretically, recently Hamner {\it et al.} made the observation of CSI experimentally in gaseous two-component BECs by accelerating the two components in opposite directions, utilizing the Zeeman shift under a magnetic-field gradient \cite{HamnerCEH2010arXiv}.
 They observed that shocks and dark--bright solitons nucleate via the instability in quasi-one-dimensional cigar-shaped two-component BECs.  
 Since such a soliton is unstable against decay into quantized vortices in two- and three-dimensional systems \cite{AndersonHRFCCA2001PRL86}, the CSI can cause vortex nucleation leading to quantum turbulence even in trapped systems with large atomic clouds \cite{TakeuchiIT2010PRL105}.

This paper presents a detailed analysis of CSI in a two-component BEC from linear instability to the nonlinear development into quantum turbulence in uniform systems, focusing on the momentum exchange between the two components and the parametric dependence of the instability. 
 In Sec. II we review the linear stability of countersuperflow in Bogoliubov analysis.
 We find that the momentum exchange is unique to the countersuperflow states.
 Sections III and IV develop the discussion into the nonlinear regime of CSI by numerically solving the Gross--Pitaevskii (GP) equations.
 Section III is devoted to density pattern formations and vortex nucleation in the early stage of the nonlinear dynamics.
 We find that the phase diagram of the linear stability characterizes the density patterns depending on the relative velocity
 and the distributions of the resulting vortex rings typically follow the structures of the patterns.
 Section IV discusses the development into vortex tangles from the vortex nucleation.
 The development of vortex lines is similar to that in the thermal counterflow of superfluid ${}^4$He.
 In our system, the momentum exchange between the two components plays the role of mutual friction between the normal fluid and superfluid components in thermal counterflow.
 Even if the total length of the initially nucleated vortex line is small,
 vortex line stretching leads to vortex tangles and the momentum exchange is completed for a large interaction between the two components.
 The vortex tangles obtained are isotropic, in contrast to the vortex tangles under thermal counterflow.
 The results are summarized in Sec. V.

\section{Countersuperflow Instability}
  Let us consider two counterpropagating superfluids consisting of two distinguishable particles in an isolated uniform system at zero temperature.
 When the relative velocity between the two superfluids exceeds a critical value,
 the countersuperflow becomes unstable.
 The instability should then induce a momentum exchange between the two superfluids to decay their relative motion.
 Note that CSI is essentially different from Landau instability \cite{KhalatnikovBook1965}, in which the superfluid system is treated in the framework of a canonical ensemble.
 In Landau instability, superfluids decay due to friction with the ``rigid'' environment, {\it e.g.}, the container wall or a normal fluid component whose velocity field is frozen to the wall due to its viscosity.
 On the other hand, CSI is purely an internal
instability with no influence from the external environment.
 Thus, the total energy of the two superfluids is conserved, keeping its total momentum constant, because of Galilean invariance.

 It is physically evident that the momentum exchange in CSI cannot initiate a reduction of the relative motion of the superfluids as a whole.
 The momentum exchange must arise from a gradual excitation of internal motions, {\it i.e.}, from the appearance of excitations in the superfluids.
 In the CSI of two-component BECs, the momentum exchange due to the excitations is analytically evaluated with the GP and the Bogoliubov--de Gennes (BdG) models.
 In this section, we discuss the linear stability and the momentum exchange in the CSI of two-component BECs in the BdG models.

 \subsection{Formulations}
 We consider a binary mixture of BECs described by the condensate wave functions $\Psi _j({\bm r},t)=\sqrt{n_j({\bm r},t)}e^{i\phi _j({\bm r},t)}$ in a mean-field approximation at $T=0$ K, where the index $j$ refers to each component ($j=1,2$).
 The wave functions are governed by the coupled GP equations \cite{Pethick}
 \begin{eqnarray}
i \hbar \frac{\partial}{\partial t} \Psi _j = \left(-\frac{\hbar^2}{2m_j}{\bm \nabla}^2+\sum_{k} g_{jk}|\Psi _k|^2\right)\Psi _j,
\label{eq:GP}
\end{eqnarray}
where $k=1$, 2 and $m_j$ is the mass of the $j$th component.
 The coefficient $g_{jk}=2\pi\hbar^2a_{jk}/m_{jk}$ represents the atomic interaction with $m_{jk}^{-1}=m_{j}^{-1}+m_{k}^{-1}$ and the s-wave scattering length $a_{jk}$ between the $j$th and $k$th components.
 Our analysis satisfies the conditions $g_{11}g_{22}>g_{12}^2$ and $g_{jj}>0$ that the two miscible condensates are stable \cite{Pethick}.

 The wave functions $\Psi_j=\Psi_j^0$ in a stationary state are written as
 \begin{eqnarray}
 \Psi _j^0=\sqrt{n_j}e^{i\left(m_{j}{\bm V}_{j}\cdot{\bm r}-\mu_jt\right)/\hbar}
 \label{eq:steady}
 \end{eqnarray}
 with the velocity ${\bm V}_j$ and the chemical potential $\mu_j$ of the $j$th component.
 The countersuperflow is realized with ${\bm V}_1 \neq {\bm V}_2$.
 We consider a collective excitation above the stationary state as $\Psi _{j}=\Psi _{j}^0+\delta\Psi _{j}$,
 where we may write the excitation wave functions $\delta\Psi _{j}$ with the usual form
 \begin{eqnarray}
 \delta \Psi _j =e^{i(m_{j}{\bm V}_{j}\cdot {\bm r}-\mu _{j} t)/\hbar}
\left\{u_{j}e^{i({\bm q}\cdot{\bm r}-\omega t)}-\left[v_{j}e^{i({\bm q}\cdot{\bm r}-\omega t)}\right]^*\right\}.
 \label{eq:delta}
 \end{eqnarray}
 By linearizing the GP quation [Eq. (\ref{eq:GP})] with respect to $\delta\Psi _{j}$, we obtain the BdG equations, whose matrix notation is
 \begin{eqnarray}
 \sigma {\cal M}{\it W}=\varepsilon{\it W},
 \label{eq:BdG}
 \end{eqnarray}
 where
 \begin{eqnarray*}
 {\cal M}=\left( 
 \begin{array}{cccc}
 h_{1}^{+} & g_{11}n_{1} & g_{12}\sqrt{n_{1}n_{2}} & g_{12}\sqrt{n_{1}n_{2}} \\
 g_{11}n_{1} & h_{1}^{-} & g_{12}\sqrt{n_{1}n_{2}} & g_{12}\sqrt{n_{1}n_{2}} \\
 g_{12}\sqrt{n_{1}n_{2}} & g_{12}\sqrt{n_{1}n_{2}} & h_{2}^{-} & g_{22}n_{2} \\
 g_{12}\sqrt{n_{1}n_{2}} & g_{12}\sqrt{n_{1}n_{2}} & g_{22}n_{2} & h_{2}^{+} \\
 \end{array}  
 \right)
 \end{eqnarray*}
 \begin{eqnarray*}
 \mbox{with}\  h_{j}^{\pm}=\epsilon _{j}^0\pm\frac{\rho_{k}}{\rho _{1}+\rho _{2}}{\bm V}_{R}\cdot\hbar{\bm q} +g_{jj}n_{j}\  (k\neq j)\  \mbox{and}
 \end{eqnarray*}
 \begin{eqnarray*}
   {\it W}=\left( 
   \begin{array}{c}
      u_{1}\\
     -v_{1}\\
      u_{2}\\
     -v_{2}\\
   \end{array} 
   \right)
   \mbox{,}\  
   \sigma =\left( 
   \begin{array}{cccc}
     1 &  0 & 0 & 0  \\
     0 & -1 & 0 & 0  \\
     0 &  0 & 1 & 0  \\
     0 &  0 & 0 & -1 \\
   \end{array} 
   \right).
 \end{eqnarray*}
 Here ${\bm V}_{R}={\bm V}_{1} - {\bm V}_{2}$ is the relative velocity, $\epsilon _{j}^0= \hbar^2{\bm q}^2/2m_{j}$, and $\varepsilon=\hbar\omega-{\bm V}_{G}\cdot\hbar{\bm q}$.
 We may neglect the center-of-mass motion of the two condensates ${\bm V}_{G}\equiv(\rho _{1}{\bm V}_{1}+\rho _{2}{\bm V}_{2})/(\rho _{1}+\rho _{2})$ with $\rho_{j}=m_{j}n_{j}$, so we set ${\bm V}_{j}$ and $\rho _{j}$ to eliminate the velocity of the center of mass.
 The eigenvector {\it W} is satisfied with
 \begin{eqnarray}
   u_{1}=-\frac{\epsilon _{1}^0-\frac{\rho _{2}}{\rho _{1}+\rho _{2}}{\bm V}_{R}\cdot\hbar{\bm q}+\varepsilon}{\epsilon _{1}^0+\frac{\rho _{2}}{\rho _{1}+\rho _{2}}{\bm V}_{R}\cdot\hbar{\bm q}-\varepsilon}v_{1}, 
      \label{eq:u1v1} \\
   u_{2}=-\frac{\epsilon _{2}^0+\frac{\rho _{1}}{\rho _{1}+\rho _{2}}{\bm V}_{R}\cdot\hbar{\bm q}+\varepsilon}{\epsilon _{2}^0-\frac{\rho _{1}}{\rho _{1}+\rho _{2}}{\bm V}_{R}\cdot\hbar{\bm q}-\varepsilon}v_{2}.
   \label{eq:u2v2} 
 \end{eqnarray}
 Because the operator $\sigma {\cal M}$ in Eq. (\ref{eq:BdG}) is non-Hermitian, the eigenvalue $\varepsilon$ can have an imaginary part. 
 When the eigenvalue has an imaginary part, the system is dynamically unstable.

 \subsection{Momentum exchange}
 We investigate how the momentum exchange happens via an excitation.
 We can define the momentum ${\bm J}_{j}$ of the $j$th component as
 \begin{eqnarray}
   {\bm J}_{j}=\int d{\bm r} \frac{\hbar}{2i}(\Psi _{j}^*\nabla\Psi _{j} - \Psi _{j}\nabla \Psi _{j}^*).
 \end{eqnarray}
 For the steady countersuperflow state, we have
 \begin{eqnarray}
   {\bm J}_{\rm total}={\bm J}_{1}+{\bm J}_{2}=N_{1} m_{1}{\bm V}_{1}+N_{2} m_{2}{\bm V}_{2},
 \end{eqnarray}
 with $N_{j}=\int d{\bm r}|\Psi _{j}|^2$.
 Here we consider that a collective excitation of Eq. (\ref{eq:delta}) appears in this state.
 The change of momentum of the $j$th component by the excitation is then
 \begin{eqnarray}
   \delta{\bm J}_{j}&=V(|u_{j}|^2 - |v_{j}|^2)\hbar{\bm q},
 \end{eqnarray}
 where $V$ is the system volume and we used the particle number conservation law of the $j$th component
 \begin{eqnarray}
  \int d{\bm r}|\Psi _{j}^{0}+\delta\Psi _{j}|^{2}=V(n_{j}+|u_{j}|^2 + |v_{j}|^2)=N_{j}.
 \end{eqnarray}
 
 In an isolated uniform system,
 the momentum conservation law requires $\delta {\bm J}_{1}+\delta {\bm J}_{2}=0$.
 Therefore, for the momentum exchange, we need the following condition to be satisfied:
 \begin{eqnarray}
 |u_{1}|^2 - |v_{1}|^2 =- |u_{2}|^2 + |v_{2}|^2 \neq 0.
 \label{eq:moex}
 \end{eqnarray}
 The condition can be evaluated using the quantity $W^\dagger{\cal M}W-(W^\dagger{\cal M}W)^*$ with ${\cal M}={\cal M}^\dagger$, which is reduced to
 \begin{eqnarray}
   (\varepsilon-\varepsilon^*)(|u_{1}|^2 - |v_{1}|^2+|u_{2}|^2 - |v_{2}|^2)=0.
 \end{eqnarray}
 When the eigenvalue $\varepsilon$ has a complex value $\varepsilon\neq\varepsilon^*$, we have $|u_{1}|^2 - |v_{1}|^2+|u_{2}|^2 - |v_{2}|^2=0$ and then the excitation is possible satisfying the momentum conservation law.

 Additionally, relative velocity is necessary to exchange momentum between the two condensates.
 The dispersion relation for ${\bm V}_{R}=0$ is well known and described by $\varepsilon^2 =\frac{1}{2}(\epsilon _{1}^2+\epsilon _{2}^2)\pm\frac{1}{2}\sqrt{(\epsilon _{1}^2-\epsilon _{2}^2)+16\epsilon _{1}^0\epsilon _{2}^0n_{1}n_{2}g_{12}^2}$ with $\epsilon _{j}^2=(\epsilon _{j}^0)^2+2\epsilon _{j}^0g_{jj}n_{j}$ \cite{Pethick}.
 Although $\varepsilon$ is purely imaginary when the interaction parameters satisfy the condition $g_{11}g_{22}<g_{12}^2$, the excitations then do not cause momentum exchange, {\it i.e.}, $\delta {\bm J}_j=0$ with $|u_j| =  |v_j|$, which is derived from Eqs. (\ref{eq:u1v1}) and (\ref{eq:u2v2}).
Accordingly, the condition (\ref{eq:moex}) is satisfied only in dynamic instability with a finite relative velocity between the two components.

 \begin{figure}
   \centering
   \includegraphics[width=1. \linewidth]{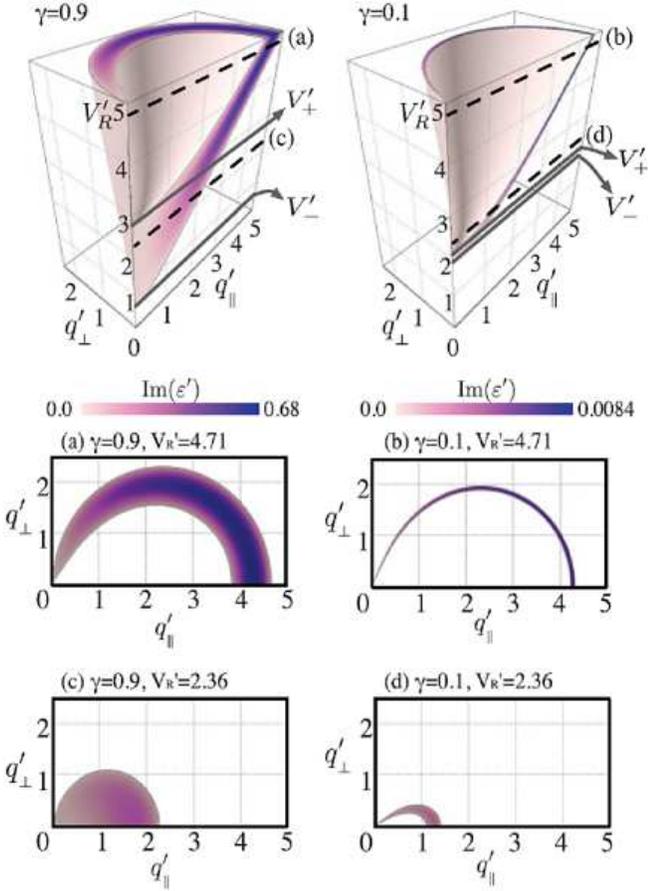}
   \caption{(Color online)
     Phase diagram of the countersuperflow instability for $\gamma=0.9$ and $0.1$.
     The gray (purple) region shows the unstable region, which shows the inequality (\ref{eq:condition}).
     The vertical axis is the relative velocity $V'_{R}=|{\bm V}'_{R}|=|{\bm V}_{R}|/c$ with $c=\sqrt{gn/m}$.
	 The horizontal axis is the wave number $q'_{\parallel,\perp}=q_{\parallel,\perp}\xi$ of the excitations, where we use $\xi=\hbar/\sqrt{mgn}$ and $q_{\parallel}'^2+q_{\perp}'^2={\bm q}'^2$ with $q'_{\parallel}=|{\bm q}'\cdot{\bm V}'_{R}/V'_{R}|$ and $q'_{\perp} \geq 0$.
     The color bars show the magnitude of the imaginary part of the eigenvalues $\varepsilon'=\varepsilon/gn$.
     The dashed lines represents $V'_{R}=4.71$ [(a) and (b)] and $2.36$ [(c) and (d)] and the solid lines show the critical velocities $V_{\pm}=2\sqrt{1\pm \gamma}$.
	 The lower four figures show the cross-section surfaces of $V'_{R}=2.36$ and $4.71$ in the phase diagrams.
   }
   \label{fig:souzu}
 \end{figure}%

 \subsection{Dispersion relation and phase diagram}
 
 We can obtain the dispersion relation by solving the eigenvalue problem of the BdG Eqs. (\ref{eq:BdG}), which yields a quartic equation in $\varepsilon$.
 Although the eigenvalue problem can in principle be solved analytically,
 the explicit expression is generally quite complicated. 
 Here, we consider the symmetric parameters with $m_{11}=m_{22}=m$, $n_{1}=n_{2}=n$, and $g_{11}=g_{22}=g$,
 which are realistic in experiments with $^{87}$Rb atoms.
 In this case, we obtain a simple form of the dispersion relation
\begin{align}
   \varepsilon'^2 =&\frac{1}{4}{\bm q}'^4+{\bm q}'^2+\frac{1}{4}q_{\parallel}'^2V_{R}'^2 \notag\\
   &\pm\sqrt{\bigl(\frac{1}{4}{\bm q}'^4+{\bm q}'^2\bigr)q_{\parallel}'^2V_{R}'^2+{\bm q}'^4\gamma^2}
   ,\label{eq:omega}
 \end{align}
 where $\varepsilon'=\varepsilon/gn$, ${\bm q}'={\bm q}\xi$ with healing length $\xi=\hbar/\sqrt{mgn}$, $V'_{R}=|{\bm V}'_{R}|=|{\bm V}_{R}|/c$ with sound velocity $c=\sqrt{gn/m}$, and $\gamma = g_{12}/g$.
 Here ${\bm q}'^2=q_{\parallel}'^2+q_{\perp}'^2$ with $q_{\parallel}'=|{\bm q}'\cdot{\bm V}'_{R}/V'_{R}|$ and $q_{\perp}' \geq 0$.
 Then, the only important parameters in the countersuperflow are $V'_{R}=V_{R}/c$ and $\gamma = g_{12}/g$.
 We have ${\rm Im}~\varepsilon' \neq 0$ if the right hand side of Eq. (\ref{eq:omega}) becomes negative.
 By comparing the square of the fourth term to that of the sum of the other three terms on the right-hand side of Eq. (\ref{eq:omega}),
 we obtain the condition of the CSI
 \begin{eqnarray}
   \sqrt{\frac{1}{4}{\bm q}'^4+{\bm q}'^2(1-\gamma)}<\frac{1}{2}q'_{\parallel}V'_{R}<\sqrt{\frac{1}{4}{\bm q}'^4+{\bm q}'^2(1+\gamma)}.
   \label{eq:condition}
 \end{eqnarray}

 Figure \ref{fig:souzu} shows this condition with $\gamma=0.9$ and $0.1$, where the gray (purple) zone shows that the countersuperflow is unstable.
 When $V'_{R}$ is small, no mode is unstable.
 However, some modes become unstable when $V'_{R}$ exceeds the critical value $V'_{-}$ with
 \begin{eqnarray}
   V'_{\pm}=2\sqrt{(1\pm\gamma)},
   \label{eq:Vc}
 \end{eqnarray}
 which is based on Eq. (\ref{eq:condition}).

 For small relative velocity with $V'_{-}<V'_{R}<V'_{+}$,
 the unstable region is broadly distributed to the lower-wave-number region [Fig. \ref{fig:souzu} (c)]. 
 When $V'_{R}$ increases to sufficiently large relative velocity with $V'_{R}>V'_{+}$,
 the unstable modes with larger values of $|{\rm Im}~\varepsilon'|$ are distributed to the higher-wave-number region in
 the wave-number space ($q'_{\parallel}, q'_{\perp}$) and the cross section forms crescents [Figs. \ref{fig:souzu}(a) and \ref{fig:souzu}(b)].


 \begin{figure*}[htb]
   \centering
   \includegraphics[width=1. \linewidth]{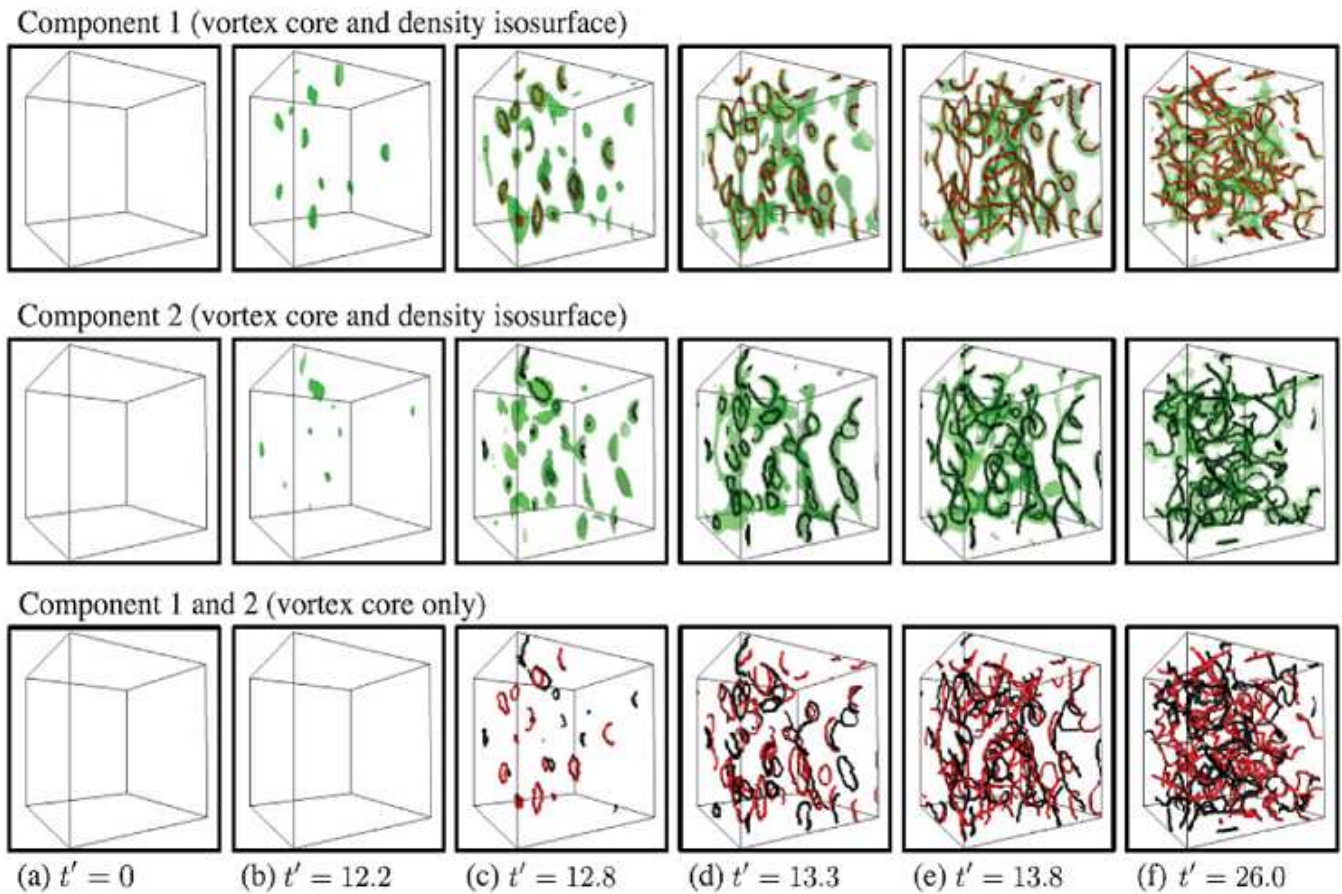}
   \caption{(Color online)
 Nonlinear dynamics of vortex cores in countersuperflow instability.
Curved lines in the top and middle panels show the vortex cores of the first and second components, respectively.
The vortex cores of both component are plotted together in the bottom panels.
The curved surfaces in top top and middle panels represent the density isosurfaces of the first component with $|\Psi_1|^2/n=0.1$ and the second component with $|\Psi_2|^2/n=0.1$, respectively.
 First, the instability causes disk-shaped low-density regions that face in the direction parallel to the initial relative velocity ${\bm V}_{R}\parallel \hat{\bm x}$ with a unit vector $\hat{\bm x}$ along the $x$ axis (b).
 Then small vortex rings appear inside the low-density disks (c) and the disk then immediately transforms into a torus-shape like a usual vortex ring (d).
 The nucleated vortex rings gradually expand due to momentum exchange between the two components.
 The vortices then start to reconnect with each other (e).
 Frequent reconnections then cause vortex tangles (f).
  The time $t$ is normalized as  $t =t\tau$ with $\tau=\hbar/gn$.
  The box size is $V=(8\xi)^3$ with $\xi=\hbar/\sqrt{mgn}$.
   }
   \label{fig:3D}
 \end{figure*}%

 \section{Vortex nucleation}

 In Sec. II we demonstrated the linear stability analysis of counterpropagating two-component BECs.
 In the linear stage of the instability, the unstable modes are amplified exponentially with time and momentum is exchanged between the two components.
  It is expected that the wave numbers of the unstable modes with larger values of $|{\rm Im}~\varepsilon'|$ could affect the density patterns emerging after the amplification in the nonlinear regime of the instability.
 Hamner {\it et al.} have observed experimentally that the CSI causes characteristic solitary waves in the nonlinear stage in a quasi-one-dimensional system of two-component BECs \cite{HamnerCEH2010arXiv}.
In a three-dimensional system, the density patterns lead to vortex nucleation in the early stage of the nonlinear development \cite{TakeuchiIT2010PRL105}.

In this section we discuss the nonlinear dynamics of CSI until vortex nucleation in a three-dimensional system by numerically solving the coupled GP equations [Eq. (\ref{eq:GP})].
 The numerical simulations were done in a three-dimensional system, which was subject to a periodic boundary condition.
 The initial state is the stationary state of Eq. (\ref{eq:steady}) with small white noise to trigger the instability \cite{noise},
 and we set the parameters as $m_1=m_2=m$, $g_{11}=g_{22}=g$, $n_1=n_2=n$, ${\bm V}'_1=-{\bm V}'_2$, and $\mu_j=mV_R^2/8+(g+g_{12})n$.
 Because the parameters of two components are symmetric, both components follow similar scenarios in the nonlinear dynamics.
 
 The vortex nucleation is generally classified into two cases, $V'_{R}>V'_{+}$ and $V'_{-}<V'_{R}<V'_{+}$, according to the configuration of the region of the unstable modes in the phase diagram (Fig. \ref{fig:souzu}).
 First, we demonstrate the vortex nucleation dynamics for a large relative velocity $V'_{R}>V'_{+}$.
 Figures \ref{fig:3D}(a)--\ref{fig:3D}(d) show the typical vortex nucleation dynamics with $V'_{R}=4.71>V'_{+}$ along the $x$ axis, $|\Psi _{j}|^2/n=1$, and $\gamma=0.9$.
 In the early stage of the nonlinear development after the exponential amplification of the unstable modes, disk-shaped low-density regions appear in both components, which face in the direction parallel to the relative velocity ${\bm V}_{R}$ [Fig. \ref{fig:3D}(b)].
 A tiny vortex ring is nucleated in the disk region [Fig. \ref{fig:3D}(c)],
 which immediately grows to a vortex ring with a certain size [Fig. \ref{fig:3D}(d)].
 Since the vortex nucleation results from the amplification of the unstable modes, decreasing the relative motion,
 the nucleated vortex rings in the $j$th component must carry the momentum in the opposite direction to the initial velocity ${\bm V}_j$, 
 where the direction of the phase gradient ${\bm \nabla} \phi_j$ inside the vortex ring is opposite that outside the ring \cite{DonnellyBook1991}.

\begin{figure}[pthb]
   \centering
   \includegraphics[width=1. \linewidth]{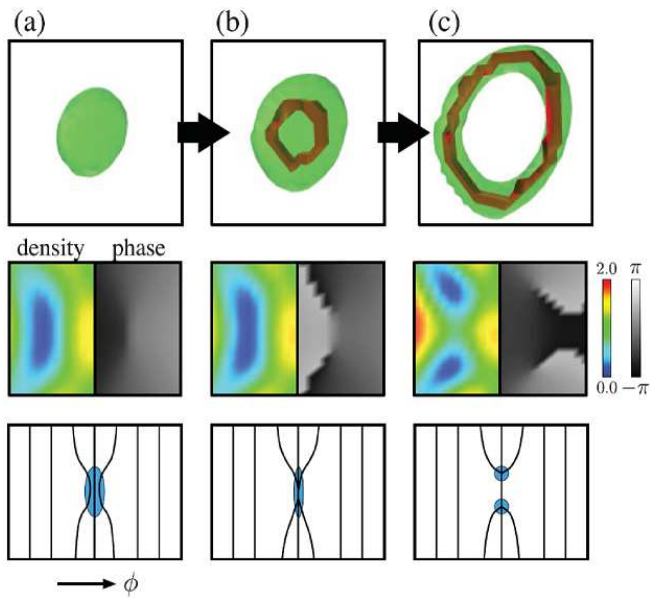}
   \caption{(Color online)
 Vortex ring nucleation from a low-density disk in countersuperflow instability.
 Top panels show the time development of the vortex core and the density isosurface of the first component with $|\Psi_1|^2/n=0.1$.
 Middle panels represents plots of the density $|\Psi_1|^2/n$ and phase $phi_1$ in the cross section cutting the ring in half.
 Shown in the bottom panels is a schematic diagram of the vortex nucleation.
 The shaded areas and solid curves show the low-density regions and the $\pi/2$-spaced isolines of phase, respectively. 
 The density patterns due to the CSI induce the disk-shaped low-density region.
 The phase gradient is steep inside the low-density disk.
 The disk decays into a vortex-antivortex pair (a vortex ring) whose size corresponds to the disk size.
 After the vortex nucleation, the phase gradient becomes smaller and the superflow is reduced between the pair (inside the ring).
   }
   \label{fig:2D}
 \end{figure}

 Figure \ref{fig:2D} shows a detailed typical dynamics of the vortex ring nucleatioin in the first component.
 First, a disk-shaped low-density region appears and the phase gradient inside the disk region becomes steep [Fig. \ref{fig:2D}(a)].
 Then a vortex ring appears from the zero-density point inside the disk region,
 where the radius of the vortex ring is smaller than that of the disk region [Fig. \ref{fig:2D}(b)].
 The radius of the ring grows immediately into that of the disk region
 and the geometry of the region transforms from a disk into a torus, like a usual vortex ring [Fig. \ref{fig:2D}(c)].
 Therefore, the size of the resulting vortex ring is roughly characterized by that of the disk-shaped low-density region.

\begin{figure}[h]
   \centering
   \includegraphics[width=1. \linewidth]{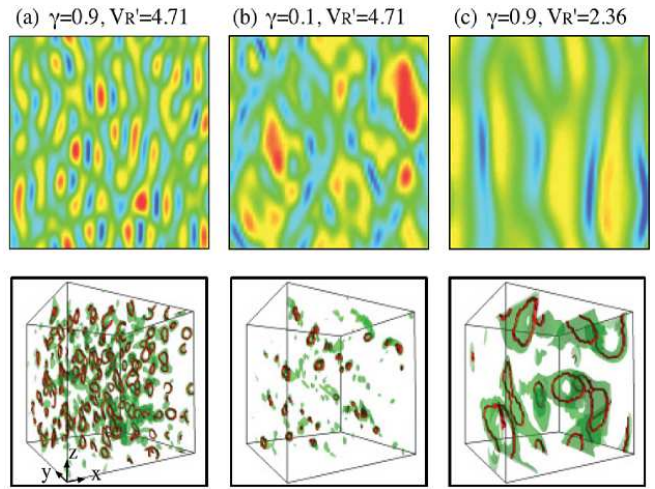}
   \caption{(Color online)
	 (Top panels) Density pattern of the first component in a cross section along the initial relative velocity.
	 (Bottom panels) Plots of cores of the nucleated vortices of the first component.
	 (a) and (b) For large relative velocity $V'_{R}>V'_+$, the low-density region is small in the direction perpendicular to the relative velocity and small vortex rings are nucleated.
	 (c) For small relative velocity $V'_+>V'_{R}>V'_-$, the density pattern is stretched in the direction perpendicular to the relative velocity and a small number of large vortex rings are nucleated.
	 The box size is $V=(16\xi)^3$.
   }
   \label{fig:VN}
 \end{figure}

 Under this consideration, we may estimate the distribution of vortex rings from the structure of the density patterns emerging in the early stage.
 The structure may be determined by the wave numbers of the unstable modes with large values of $|{\rm Im}~\varepsilon'|$ in the phase diagram (Fig. \ref{fig:souzu}).
 In the limit of large relative velocity $V_R' \gg V'_+$,
 the unstable region described with the inequality in Eq. (\ref{eq:condition}) is reduced to the circle
 \begin{eqnarray}
   (q'_{\parallel}-\frac{1}{2}V'_{R})^2+q_{\perp}'^2=\frac{1}{4}V_{R}'^2.
   \label{eq:circle}
 \end{eqnarray}
  As a result, the wave numbers of the unstable modes are characterized by $q'_{\parallel} \lesssim V_{R}'$ and $q'_{\perp} \lesssim V_{R}'/2$ for sufficiently large relative velocity.
   Figure \ref{fig:VN}(a) shows a typical density pattern in a cross-section plane along the relative velocity (top panels) and its vortex plot (bottom panels).
 Most of the low-density disks nucleate vortex rings. 
 Thus the periodicities $\sim V_R/\kappa$ of the density pattern characterize the vortex ring size and the interval between the vortex rings, where we used $\kappa\equiv 2\pi\hbar/m$.
 Therefore, for $V_R' > V'_+$, the vortex line density just after the nucleation event may be roughly estimated to be $\sim V_{R}^2/\kappa^2$, which is independent of $\gamma$.

However, the situation is different when $\gamma$ or $V_{R}$ becomes small.
 When the interaction between the two components is weak for small $\gamma$,
 the density pattern is highly disturbed by the time the vortex rings are nucleated [Fig. \ref{fig:VN}(b) (top)].
 As a result, the number of nucleated rings in Fig. \ref{fig:VN}(b) is smaller than that in Fig. \ref{fig:VN}(a),
 although the rings have the same characteristic size $\sim V_R/\kappa$.

 If the relative velocity is small, {\it e.g.}, $V'_{-}<V'_{R}<V'_{+}$,
 since the unstable modes with large $|{\rm Im}~\varepsilon'|$ are distributed to low values of $q'_{\perp}$ [Fig. \ref{fig:souzu}(c)],
 the periodicity of the density pattern perpendicular to the relative velocity becomes larger than that parallel to the velocity [Fig. \ref{fig:VN}(c) (top)].
 Thus the radius of nucleated vortex rings is larger than the interval between the rings along the relative velocity [Fig. \ref{fig:VN}(c) (bottom)],
 where the vortex line density becomes small compared to $V_{R}^2/\kappa^2$.
 If the perpendicular periodicity of the density pattern becomes comparable to the system size,
 the CSI does not make vortex rings, but rather makes vortex lines across the system or distorted stripe patterns, as observed by Hamner {\it et al.} \cite{HamnerCEH2010arXiv}. 

 \section{Transition to vortex tangle}
 Since the momentum transported by a vortex ring increases as its radius increases \cite{DonnellyBook1991},
  further momentum exchange expands the nucleated vortex rings without changing their direction.
 As the vortex rings expand, the intervortex interaction between neighboring vortex rings becomes important, leading to highly nonlinear dynamics.

 The typical dynamics after vortex nucleation are shown in Figs. \ref{fig:3D}(d)--\ref{fig:3D}(f).
 At the beginning of the expansion, the size of the nucleated vortex rings gradually becomes large without changing their direction [Fig. \ref{fig:3D}(d)].
 As the vortex rings expand further,
 they are more distorted by the interaction and reconnection with neighboring rings [Fig. \ref{fig:3D}(e)].
 The vortex distortions are more affected by the vortices in the same component than by those in the different components.
This is because the force between vortices in the same component is stronger than that in the different components for large distances \cite{arXiv_Eto}.
 In addition, vortices in the same component can make the usual reconnections \cite{DonnellyBook1991},
 while those in the different components just cross each other even if they collide.
 The distortion leads to a tangle of quantized vortices
 and then both components follow the same sequence to form binary quantum turbulence [Fig. \ref{fig:3D}(f)]

This scenario is similar to that of the transition to quantum turbulence in thermal counterflow \cite{DonnellyBook1991}.
 In the transition dynamics to thermal counterflow turbulence,
the vortex lines attached to the container wall are dragged and stretched by mutual friction,
 which causes a momentum exchange between the superfluid and normal fluid components.
 In our system of CSI, the momentum exchange between the two components plays the role of mutual friction in the thermal counterflow.
 The mutual friction is microscopically understood as the reaction of
 scattering of the excitations, of which the normal fluid component consists, at the core of the quantized vortices.
 However, this picture cannot be simply applied to the ``mutual friction'' in two-component BECs because neither of the two components is a normal component.
 A microscopic understanding of the ``mutual friction'' is an open problem for future work.

 There are two evident differences between CSI and thermal counterflow instability.
 One difference is the vortex nucleation mechanism.
 In CSI,
 the vortices are {\it intrinsically} nucleated in the bulk, as discussed above,
 whereas in thermal counterflow the vortices are prepared on the boundary of the system before the instability.
 The second difference involves the relative motion between the two components.
 Thermal counterflow turbulence has been traditionally studied in states in which the superfluid component is turbulent but the normal fluid component is laminar sustained by an externally driving temperature gradient.
 For CSI, in contrast,
 the relative motion decays through the momentum exchange between the two components and both components become turbulent, keeping its total momentum conserved in an isolated uniform system \cite{ref_GuoCNVM2010PRL105}.

\begin{figure}
   \centering
   \includegraphics[width=1. \linewidth]{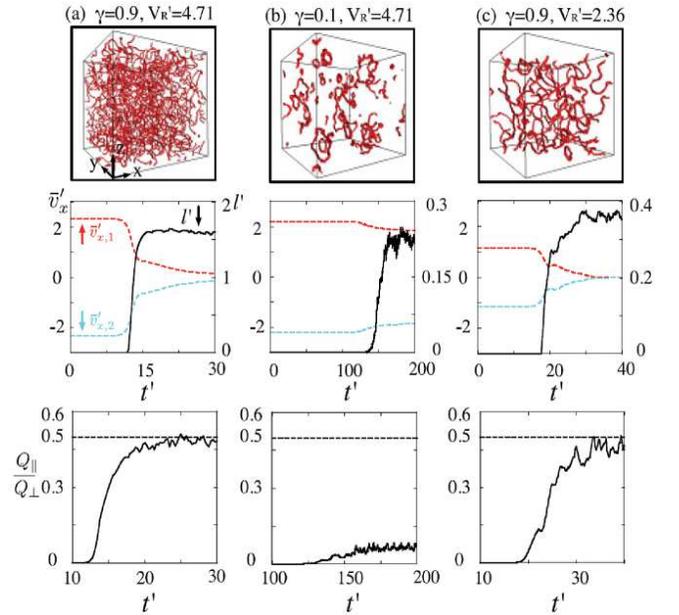}
   \caption{(Color online)
     (Top panels) Plots of vortex cores of the first component at the time when the vortex line density is maximum.
 The box size is $V=(16\xi)^3$.
     (Middle panels) Time evolution of the total vortex line density $l'=l\xi^2$ of both components and the normalized mean superfluid velocity $\bar{v}'_{x, j}\xi/\tau=J_{x, j}/m_jN_j$ with the momentum $J_{x,j}$ of the $j$th component in the $x$ direction. 
	 (Bottom panels) Time evolution of the parameter $Q_{\parallel}/Q_{\bot}$.
	 The momentum exchange happens drastically in the expansion of the nucleated vortex rings.
	 The rate of the momentum exchange is reduced when vortex reconnection occurs frequently in the vortex tangles.
	 Eventually, the momentum exchange is completed, leading to an isotropic tangle.
	 For small $\gamma=0.1$ (b), the vortex rings expand less and thus the momentum exchange is completed without an isotropic tangle.
	 Even if the vortex line density just after the vortex nucleation event is small for small relative velocity $V'_+>V'_R>V'_-$ with large $\gamma=0.9$,  the vortices are well stretched and repeat reconnections to develop into a dilute isotropic tangle, where the momentum exchange is also completed.
   }
   \label{fig:VE}
 \end{figure}

 In order to obtain further insight into CSI,
 we investigate how the relative motion decays, stretching vortices throughout the nonlinear dynamics.
 The typical time development of the mean superfluid velocity $\bar{v}_{x,j}=J_{x,j}/m_jN_j$ with the momentum $J_{x,j}$ of the $j$th component in the $x$ direction is plotted together with the total vortex line density $l$ in Fig. \ref{fig:VE}(a) (middle).
 The rate of the momentum exchange drastically increases after the vortex nucleation
 and decreases when vortex reconnection takes place frequently.
 This implies that the expansion of vortex rings is the most efficient mechanism for momentum exchange in CSI.
 After vortex tangles are realized via frequent vortex reconnection,
 the momentum exchange moderately continues even when the averaged relative velocity becomes lower than the critical relative velocity, {\it i.e.}, $|\bar{v}_{x,1}-\bar{v}_{x,2}|<V_-$,
 and then the momentum exchange is almost completed and the relative motion vanishes globally.

 In the momentum exchange process,
 the kinetic energy due to the {\it global} relative motion with $\bar{v}_{x,1} \neq \bar{v}_{x,2}$ is mainly consumed in driving the {\it internal} motion, {\it e.g.}, nucleation and stretching of vortices in each component.
 This is why the growth rate of the vortex line density follows the rate of momentum exchange in Fig. \ref{fig:VE}(a) (middle),
 {\it e.g.}, the growth rate of the vortex line decreases when the rate of momentum exchange becomes small after frequent vortex reconnection.
 Generally speaking, the vortex line length is reduced by emitting phonons via vortex reconnection \cite{DonnellyBook1991}
 and hence the vortex line density eventually stops growing when the reduction due to vortex reconnection exceeds the growth effect by momentum exchange.

\begin{figure}
   \centering
   \includegraphics[width=1. \linewidth]{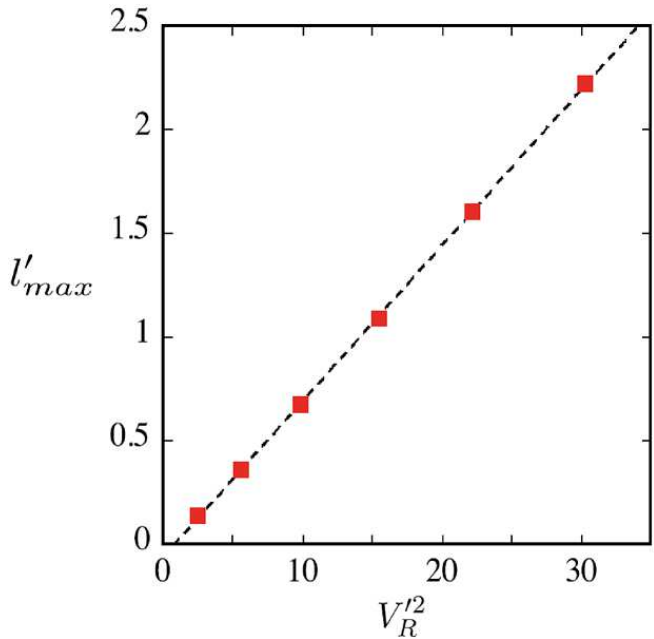}
   \caption{(Color online)
   Maximum total vortex line density $l'_{\max}=l_{\max}\xi^2$ in the CSI turbulence with $\gamma=0.9$ as a function of $V_{R}'^2$,
 which is obtained by making an ensemble average of five numerical results with different initial noise.
   }
   \label{fig:VLD}
 \end{figure}

 The maximum value $l_{\max}$ of the vortex line density should increase with the initial relative velocity $V_{R}$ since the initial kinetic energy is consumed in stretching vortex lines.
 Figure \ref{fig:VLD} shows the numerical result of the maximum vortex line density $l_{\max}$ dependence on $V_R$ for $\gamma=0.9$.
 The vortex line density $l_{\max}$ is proportional to $V_R^2$.
 This behavior of $l_{\max}$ is phenomenologically explained from the energetics considering the tension of a vortex line.
 To apply this consideration to the CSI turbulence,
 we assume that all of the kinetic energy due to the global relative motion is used for stretching vortices.
 Then the kinetic-energy density $\nu_0$ in the initial state is equal to $\alpha l_{\max}$,
 where the proportionality factor $\alpha$ represents the effective tension coefficient of the vortex line.
 Since the energy density $\nu_0$ is written as $\nu_0 = \frac{\rho_1}{2}V_1^2+\frac{\rho_2}{2}V_2^2=\frac{\varrho}{2}V_R^2$ with $V_G=0$ and $\varrho=\frac{\rho_1\rho_2}{\rho_1+\rho_2}=mn/2$,
 we obtain 
 \begin{eqnarray}
 l_{\max} \propto \frac{V_R^2}{\kappa^2},
 \label{eq:l_max}
 \end{eqnarray}
 where we have used $\alpha \propto n\hbar^2/m$ as an analog to the tension of an isolated vortex line.
 The numerical result of Fig. \ref{fig:VLD} may imply that $\alpha$ is not sensitive to $l$, 
 although, in general, $\alpha$ depends logarithmically on the value $a/\xi_v$ with the mean intervortex distance $a=l^{-1/2}$ and the vortex core radius $\xi_v$ \cite{DonnellyBook1991}.

 The usual thermal counterflow turbulence is anisotropic in steady states because relative motion remains between the superfluid and normal fluid components, sustained by an externally driving temperature gradient \cite{DonnellyBook1991}.
 It is expected that CSI turbulence can be isotropic since the relative motion vanishes in the final stage. 
 We investigate the anisotropy of CSI turbulence in a manner similar to the analysis of thermal counterflow turbulence by introducing the anisotropic parameters \cite{TakeuchiIT2010PRL105}
 \begin{eqnarray*}
Q_{\parallel}\equiv \frac{1}{2V}\int \omega_x^2 d{\bm r},\  Q_{\bot}\equiv\frac{1}{2V}\int (\omega_y^2+\omega_z^2)d{\bm r},
 \end{eqnarray*}
 where we have used the system volume $V$ and the effective vorticity ${\bm \omega}=(\omega_x,~\omega_y,~\omega_z)^T = {\bm \nabla}\times{\bm v}$ with the total mass-current velocity ${\bm v}({\bm r})=(\rho _{1}{\bm v}_1+\rho _{2}{\bm v}_2)/(\rho _{1}+\rho _{2})$ and ${\bm v}_{j}({\bm r})=\hbar/m_{j}{\bm \nabla} \phi_{j}({\bm r})$.
 Analogous to thermal counterflow turbulence,
 we have $Q_{\parallel}/Q_{\bot}=1/2$ for isotropic CSI turbulence. 
 Conversely, if the tangle consists of curves lying only in planes perpendicular to ${\bm V}_{R}$, then we have $Q_{\parallel}/Q_{\bot}=0$.

 We show the typical time development of $Q_{\parallel}/Q_{\bot}$ from the vortex nucleation in Fig. \ref{fig:VE} (a) (bottom).
 $Q_{\parallel}/Q_{\bot}$ is almost zero just after the vortex nucleation because the vorticity ${\bm \omega}$ of the nucleated vortex ring is perpendicular to the relative velocity.
 After the vortex rings start to interact with other vortices and get distorted,
 the vortex tangle grows to be isotropic $Q_{\parallel}/Q_{\bot} \sim 1/2$ as the momentum exchange is completed $\bar{v}_{x,1}=\bar{v}_{x,2}=0$.

The above analysis is not applicable to the smaller $\gamma$ case, where the nucleated vortex rings expand little and reconnect less with the other vortex rings [Fig. \ref{fig:VE} (b)].
 In this case, vortex lines tend to lie perpendicular to the relative velocity with $Q_{\parallel}/Q_{\bot} \ll 1/2$,
 where there remain vortex rings facing parallel to the relative velocity, as seen in Fig. \ref{fig:VE} (b) (top).
 Then the momentum exchange is not completed and slows even when the averaged relative velocity is still much larger than the critical relative velocity: $|\bar{v}_{x,1}-\bar{v}_{x,2}|>V_-$.

 In contrast, the relative velocity has less influence on the nonlinear dynamics.
 For small $V_{R}$ with $\gamma=0.9$, the vortex rings expand readily and repeat reconnections to develop into isotropic turbulence without global relative motion of the two components [Fig. \ref{fig:VE} (c)].
 It is interesting that
 the vortex tangles are anisotropic in Fig. \ref{fig:VE} (b) but isotropic in Fig. \ref{fig:VE} (c),
 although the values of the maximum vortex line density are comparable for both cases.
 This shows that large $\gamma\ (<1)$ rather than large $V_R$ is crucial for realizing an isotropic tangle.

\section{Summary}
 We studied theoretically the instability of countersuperflow in uniform miscible two-component BECs.
 It was shown that the dynamic instability induces momentum exchange between the two components with a finite relative velocity in the Bogoliubov analysis.
 The nonlinear development of the instability was analyzed by numerically solving the coupled GP equations.
 The distribution of the nucleated vortex rings was qualitatively evaluated from the instability phase diagram.
 The size of the nucleated vortex rings increases as the initial relative velocity $V_R$ decreases.
 The nonlinear dynamics after vortex nucleation are similar to those of the turbulence transition in thermal counterflow of superfluid ${}^4$He.
 In contrast to the anisotropy of thermal counterflow turbulence,
 CSI causes isotropic turbulence by completing the momentum exchange.
 The maximum vortex line density of the turbulence is proportional to $(V_R/\kappa)^2$.
 Even if the total line length of the nucleated vortices is small for small $V_R$,
 isotropic turbulence is realized by vortex stretching due to mutual friction for a sufficiently large intercomponent interaction.

 In this work we considered uniform density condensates.
 However, density is inhomogeneous in the experimental realizations of BEC in dilute atomic clouds.
 Our results may still be applied to the case in which the size of an atomic cloud is large enough compared to the density pattern periodicity $\sim V_R/\kappa$
 parallel and perpendicular to the relative velocity ${\bm V}_R$ with large $V_R>V_+$.
 Since the critical relative velocity decreases with the condensate density,
 the critical relative velocity at the surface of the condensates is smaller than that at the center in the local density approximation.
 Thus the instability can occur from the surface before the relative velocity exceeds the critical relative velocity at the center.
 Then vortices would be nucleated not in the bulk but from the surface after the amplification of the unstable surface modes.

\begin{acknowledgements}
 We acknowledge H. Adachi for useful discussions.
 Part of the calculation was done using the facilities of the Supercomputer Center at the Institute for Solid State Physics, University of Tokyo.
 This work was supported by the "Topological Quantum Phenomena" (No. 22103003) Grant-in Aid for Scientific Research on Innovative Areas from the Ministry of Education, Culture, Sports, Science and Technology (MEXT) of Japan.
M.T. acknowledges the support of a Grant-in-Aid for Scientific Research from JSPS (Grant No. 21340104).
 \end{acknowledgements}



\begin{thebibliography}{99}
\bibitem{noise} When the exponential growth rates ${\rm Im}~\varepsilon'$ of the unstable modes are high for large relative velocity,
 the nonlinear dynamics is not so strongly dependent on the small thermal noise at very low temperatures.
\end{thebibliography}
\end{document}